\pdfoutput=1
\documentclass[10pt,aps,prb,amsmath,amssymb,twocolumn,%
                letterpaper,footinbib,showpacs,raggedbottom,%
                citeautoscript,balancelastpage,%
                floatfix]{revtex4-1}

\usepackage[usenames,dvipsnames]{xcolor}
\usepackage[bookmarks=false,colorlinks]{hyperref}
\hypersetup{
    linkcolor=magenta,        
    citecolor=PineGreen,     
    filecolor=Plum,         
    urlcolor=blue,           
}
\usepackage{graphicx}
\usepackage{mdwlist}
\usepackage[protrusion=true,expansion=true,final]{microtype}



\begin{document}

\title{Improper Ferroelectricity and Piezoelectric Responses in Rhombohedral ($A$,$A^{\prime}$)$B_2$O$_6$ Perovskite Oxides}
\author{Joshua Young}
	\email{jy346@drexel.edu}
	\affiliation{Department of Materials Science \& Engineering,\! Drexel University,\! Philadelphia,\! PA 19104,\! USA}
\author{James M.\ Rondinelli}
	\email{jrondinelli@coe.drexel.edu}
	\affiliation{Department of Materials Science \& Engineering,\! Drexel University,\! Philadelphia,\! PA 19104,\! USA}
\date{\today}

\begin{abstract}
High-temperature electronic materials are in constant demand 
as the required operational range for various industries increases.
Here we 
design ($A$,$A^\prime$)$B_2$O$_6$ perovskite oxides with [111] ``rock salt" $A$-site cation order and predict them to be potential high-temperature piezoelectric materials. 
By selecting bulk perovskites which have a tendency towards only out-of-phase $B$O$_6$ rotations, we avoid possible staggered ferroelectric to paraelectric phase transitions while also retaining non-centrosymmetric crystal structures necessary for ferro- and piezoelectricity.
Using density functional theory calculations, we show that (La,Pr)Al$_2$O$_6$ and (Ce,Pr)Al$_2$O$_6$ display spontaneous polarizations in their polar ground state structures; we also compute the dielectric and piezoelectric constants for each phase.
Additionally, we predict the critical phase transition temperatures for each material from first-principles, to 
demonstrate that the piezoelectric responses, which are comparable to traditional lead-free piezoelectrics, should persist to high temperature.
These features make the rock salt $A$-site ordered aluminates candidates for high-temperature sensors, actuators, or other electronic devices.
\end{abstract}



\maketitle

\section{Introduction}

Ferroelectric materials are of great scientific and technological importance, especially for use in electronic devices such as non-volatile memory \cite{Scott:Book,Auciello/Scott/Ramesh:1998}, tunable capacitors \cite{Defay:2013,Jamil_Kalkur:2007}, and tunnel junctions.\cite{Shuravlev_etal:2005}
The fact that there are so few mechanisms capable of generating spontaneous polarizations with high Curie temperatures in crystalline systems, however, suggests that there are limited opportunities to improve upon existing devices.
Understanding how ferroelectricity arises, as well as how to control and purposefully induce it in novel ways, is therefore critical to engineering new technologies based on ferroic transitions.

The family of $AB$O$_3$ perovskite oxides is ideal for investigations of this nature, as the crystal family can accommodate a wide range of chemistries and exhibits strong electron-lattice coupling, which  allow for highly tunable ground states through atomic substitution and strain.\cite{Wolfram/Ellialtioglu:Book,Dawber/Rabe/Scott:2005,Pena/Fierro:2001}
Typically, ferroelectricity in transition metal perovskite oxides is realized by having a $d^0$ $B$-site cation or lone-pair active $A$-site cation, \cite{Burdett:1981,Bersuker:2001} which undergo collective polar displacements  through the so-called second order Jahn-Teller effect \cite{Pearson:1983,Halasyamani/Poeppelmeier:1998}. 
An alternative strategy receiving considerable interest lately relies on an improper ``geometric route,'' whereby a ferroelectric polarization is produced as a by-product of some complex lattice distortion(s), as opposed to being directly driven by inversion lifting displacements originating in  the chemistries of the cations.\cite{Levanyuk:1974,VanAken/Spaldin-2004}
Electric polarizations can arise in this manner through, \textit{e.g.}, rotations of corner-connected $B$O$_6$ octahedra that are coupled to and induce polar displacements of the $A$ and $B$ cations in perovskites.
If the compound is ordered as ($A$,$A^\prime$)$B_2$O$_6$, then two chemically distinct $A$-sites will displace by different amounts due to a size differential, which leads to an observed electric dipole.\cite{Benedek12}
Previous efforts have defined the octahedral rotational patterns which, when combined with alternating $A$ and $A^{\prime}$ atoms along the [001] crystallographic axis, lift the symmetries preventing spontaneous polarizations. 
In the case of these `layered' superlattices, only a combination of in-phase (+) and out-of-phase (--) $B$O$_6$ octahedral rotations ($a^+b^-b^-$ in Glazer notation) satisfies the required criteria put forth. \cite{Rondinelli/Fennie:2012}
Energetically, this occurs because the modes describing the in-phase and out-of-phase rotations (identified by the irreducible representations $M_3^+$ and $R_4^+$ of $Pm\bar{3}m$, respectively) couple to a polar mode,\cite{Mulder/Rondinelli/Fennie:2013} resulting in an anharmonic term in the free energy expansion  that ultimately stabilizes the polar structure.
Materials with spontaneous polarizations resulting from this coupling of two  zone-boundary modes are known as ``hybrid improper" ferroelectrics.\cite{Benedek/Fennie:2011}

\begin{figure}
\centering
\includegraphics[width=0.95\columnwidth,clip]{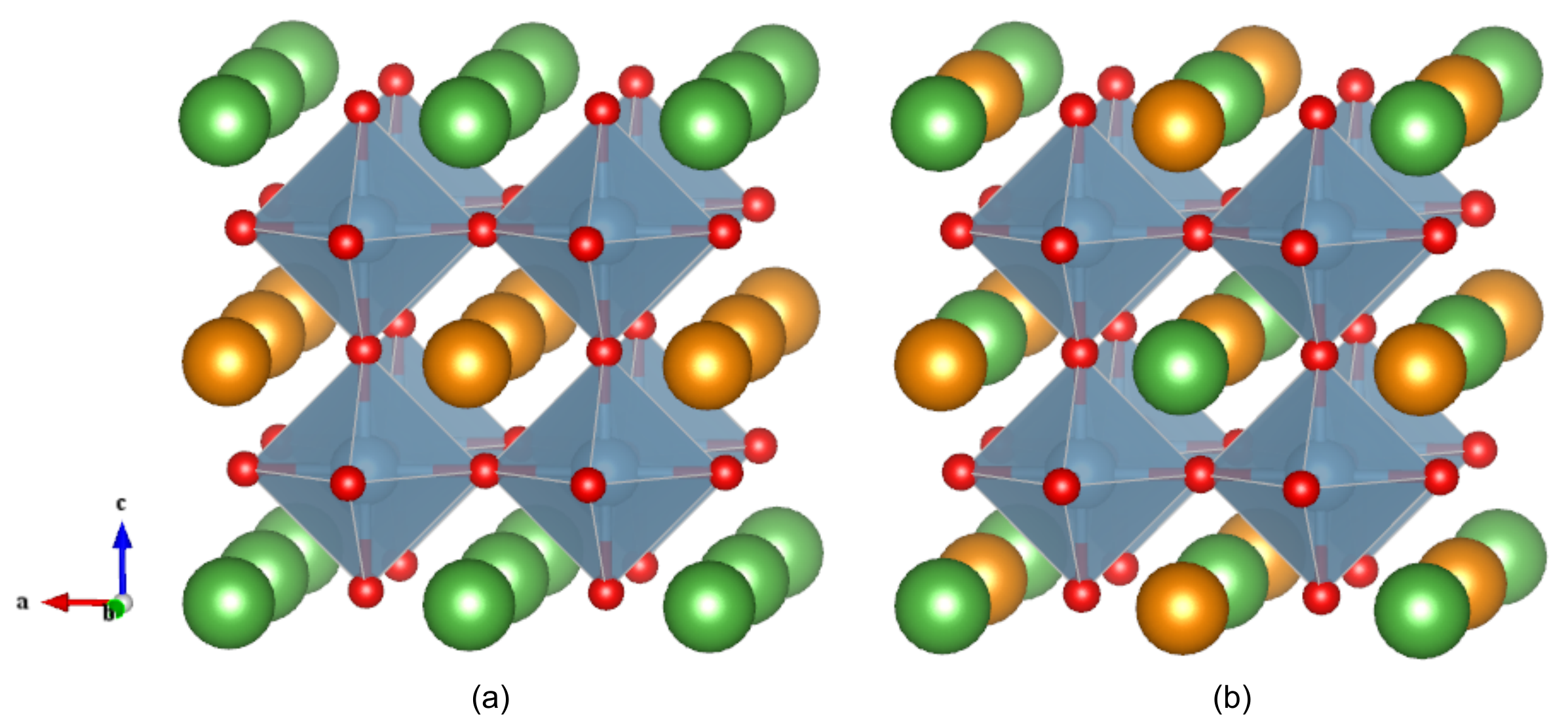}\vspace{-1\baselineskip}
\caption{An ($A$,$A^{\prime}$)$B_2$O$_6$ superlattice ordered along the (a) [001]-direction (`layered') and (b) [111]-direction (`rock salt'). $A$  and $A^\prime$ atoms are in green and orange, respectively, while $B$O$_6$ octahedra are given in blue.
}
\label{fig:orderings}
\end{figure}

The fact that [001]-ordering is conducive to layer-by-layer growth of double perovskites through pulsed laser deposition or molecular beam epitaxy,\cite{Zubko/Triscone_et_al:2011} and that the required $a^+b^-b^-$ tilt pattern is one of the most commonly adopted by perovskite oxides, make these compounds attractive for investigation.
However, the fact that the polarization arises from a trilinear coupling leads to an often overlooked problem: mode condensation.
There is the real possibility that the two rotational modes condense at different temperatures, leading to a \emph{staggered} paraelectric to ferroelectric phase transition.
Although an avalanche transition (in which the modes condense at the same temperature) is possible in ferroelectric oxides \cite{Etxebarria:2010}, non-avalanche transitions are also likely to occur. \cite{Machado:2008}
This makes it difficult to predict whether or not there will be two phase transitions from the paraelectric to ferroelectric state, \emph{i.e.}, a cell-doubling antidistortive transition, owing to the rotations, followed by an inversion lifting one, in the material without detailed and careful experimental study.
However, it is possible to avoid this problem if the spontaneous polarization is induced by only a single rotational mode, making such materials improper rather than hybrid-improper ferroelectrics.
Materials of this nature could find use in high-temperature electronic applications such as sensors.\cite{Shaulov:1980} 
Our previous work has shown that although both in-phase and out-of-phase rotations are required to lift inversion symmetry in [001]-ordered double perovskites, only out-of-phase rotations, which are described by a single mode, are needed if the $A$-sites are ordered along the [111] direction (the difference in cation ordering is shown in \autoref{fig:orderings}). \cite{Young/Rondinelli:2013}
In this work, we examine the feasibility of this symmetry result using density functional theory calculations, and subsequently design a series of [111]-ordered $(A,A^{\prime})B_2$O$_6$ double perovskites exhibiting only out-of-phase rotations with modest electronic polarizations.
We also investigate their piezoelectric response and estimate the phase transition temperatures based on first-principles total energy differences. 
We find that while two compounds, (La,Pr)Al$_2$O$_6$ and (Ce,Pr)Al$_2$O$_6$, exhibit spontaneous polarizations in their polar ground state, they undergo a transition to a chiral phase at low temperatures accompanied by a disappearance of the polarization; albeit, they then exhibit an order of magnitude increase in the piezoelectric response. 
These materials then transform to a centrosymmetric phase at 1300 K.
While the third compound, (La,Nd)Al$_2$O$_6$, does not have a polar ground state, we predict it remains non-centrosymmetric to even higher temperatures.

\section{Computational Methods}

All investigations were performed using density functional theory\cite{Hohenberg/Kohn:1964} as implemented in the Vienna \textit{ab-initio} Simulation Package (\texttt{VASP}).\cite{Kresse/Hafner:1993,Kresse/Hafner:1996}
We used projector augmented-wave (PAW) potentials\cite{Blochl:1994} with the PBEsol functional\cite{PBEsol:2008}, with the $5s^25p^65d^16s^2$ valence electron configuration for La , $5s^25p^64f^16s^2$ for Nd, $5s^25p^66s^25d^1$ for Pr, $5s^25p^65d^16s^2$ for Ce, $3s^23p^1$ for Al, and $2s^22p^6$ for O.
A plane-wave cutoff of 550 eV and a 6$\times$6$\times$6 Monkhorst-Pack mesh\cite{Monkhorst/Pack:1976} were used during the structural relaxations.
The phonon band structure of each compound was computed using density functional perturbation theory\cite{Baroni/deGironcoli/DalCorso:2001} with an increased plane-wave cutoff of 800 eV.
The ground states were determined by performing a full structural relaxation on 15 candidate structures, which were generated using linear combinations of unstable phonons of each compound.
The electric polarization was calculated using the Berry phase method\cite{King-Smith/Vanderbilt:1993,RevModPhys.66.899} as implemented in \texttt{VASP}.
Finally, the piezoelectric and elastic tensors were computed within density functional perturbation theory.\cite{Piezo_DFPT_2,Piezo_DFPT_1}

We note that DFT calculations on Ce-based compound often treat the $4f$ electron explicitly, while also including a Hubbard-$U$ correction of 6-10\,eV 
on those states.
Here we use a Ce PAW which does not include the $4f$ states, instead placing these electrons in the core to eliminate any adjustable $U$  parameters in our calculations.
We evaluated this approximation by performing a structural relaxation of bulk CeAlO$_3$, treating the $f$-electron both in the core  
and as valence electrons with $U=0$-10\,eV in increments of 1\,eV.
The relaxed lattice parameters and rotation angles of CeAlO$_3$ with the $f$-electrons in the core and treated explicitly as valence electrons with $U=8$ both agree with the experimental data to within 1\%, showing that there is essentially no error if the $f$-electrons are not included.
Owing to the structural agreement, all presented results use the former PAW. 

\section{Results and Discussion}
\subsection{Bulk Rare Earth Aluminate Ground States}

We first determined the ground state structures of four non-polar aluminate dielectrics which contain only out-of-phase AlO$_6$ octahedral rotations: LaAlO$_3$, NdAlO$_3$, PrAlO$_3$, and CeAlO$_3$.
The computationally and experimentally determined space group, lattice constants, and octahedral rotation angles and tilt patterns are summarized in \autoref{tab:bulk}.
We find that LaAlO$_3$ and NdAlO$_3$ both exhibit the $a^-a^-a^-$ tilt pattern (space group $R\bar{3}c$), in good agreement with experimental results. \cite{LAO_structure:2000, Aluminates_structure:2000}
Next, we find that PrAlO$_3$ exhibits the $a^0b^-b^-$ tilt pattern (space group $Imma$) in its ground state.
Although this compound is found to exhibit space group $C2/m$ with an $a^0b^-c^-$ tilt pattern at low temperatures experimentally, we find this phase is 10 meV per formula unit (f.u.) higher in energy than the $Imma$ phase; however, PrAlO$_3$ is known to undergo a continuous phase transition to $Imma$ at 150 K. \cite{PAO_structure:2001}
Finally, we find that CeAlO$_3$ displays the $a^0a^0c^-$ tilt pattern (space group $I4/mcm$), which is consistent with the experimentally known low-temperature structure.\cite{Fu_Ijdo:2006}

\begingroup
\squeezetable
\begin{table}[t]
\begin{ruledtabular}
\caption{\label{tab:bulk}
Structural details of the bulk aluminate superlattice constituents.  Each has a centrosymmetric space group (S.G.) and only out-of-phase rotations about one-, two-, or three-axes;  the magnitude of which is given by $\Theta$ and determined by measuring the Al-O-Al bond angle $\theta$, $\Theta = (180^{\circ}-\theta)/2$. The tolerance factor\cite{Goldschmidt:1926} ($\tau$) 
is computed using bond lengths obtained from the bond valence model.\cite{Lufaso/Woodward:2001} The calculated lattice parameters (in \AA) and rotational angle (in degrees) are compared to experimental (Exp.) data (\emph{cf}.\ text for references).} 
\begin{tabular}{lccccc}
Compound & S.G. & Tilt & $\tau$ & Theory & Exp.\ \\
\hline
LaAlO$_3$ & $R\bar{3}c$ & $a^-a^-a^-$ & 0.995 & $a$ = 5.357 & $a$ = 5.359 \\
 & & & & $c$ = 13.22 & c = 13.08 \\
 & & & & $\Theta = 5.8$ & $\Theta = 5.5$ \\
 \hline
NdAlO$_3$ & $R\bar{3}c$ & $a^-a^-a^-$ & 0.975 & $a$ = 5.325 & $a$ = 5.333 \\
 & & & & $c$ = 12.85 & c = 12.98 \\
 & & & & $\Theta = 8.4$ & $\Theta = 9.4$ \\
 \hline
PrAlO$_3$ & $Imma$ & $a^0b^-b^-$ & 0.982 & $a$ = 5.351 & $a$ = 5.339 \\
 & & & & $b$ = 7.504 & $b$ = 7.494 \\
 & & & & $c$ = 5.303 & $c$ = 5.291 \\
 & & & & $\Theta = 8.9$ & $\Theta = 9.4$ \\
 \hline
CeAlO$_3$ & $I4/mcm$ & $a^0a^0c^-$ & 0.988 & $a$ = 5.326 & $a$ = 5.309 \\
 & & & & $c$ = 7.591 & $c$ = 7.599 \\
 & & & & $\Theta = 7.2$ & $\Theta = 8.1$ \\
\end{tabular}
\end{ruledtabular}
\end{table}
\endgroup

\subsection{$A$-Cation Ordered Ground States Structures}

\begin{figure*}
\centering
\includegraphics[width=1.90\columnwidth]{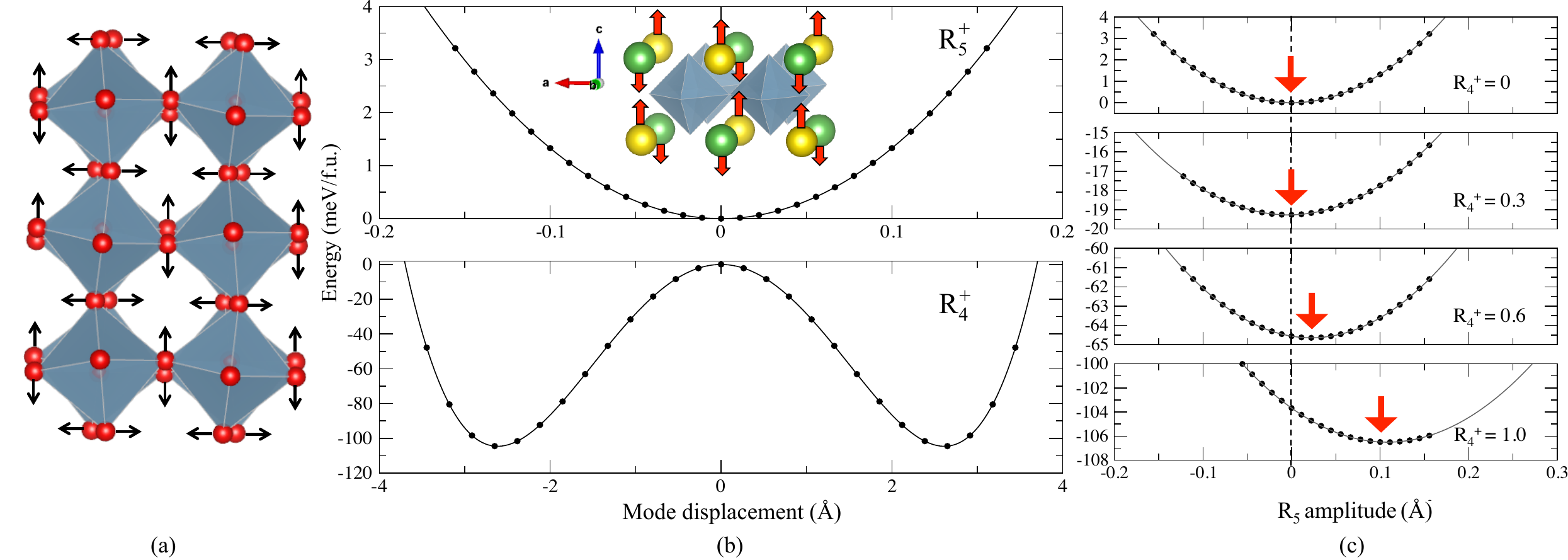}\vspace{-0.8\baselineskip}
\caption{(a) The atomic displacements constituting the out-of-phase octahedral rotations described by the $R_4^+$ irreducible representation in (La,Pr)Al$_2$O$_6$ and (Ce,Pr)Al$_2$O$_6$. (b) Energy as a function of atomic displacements (inset) described by the $R_5^+$ (upper) and $R_4^+$ (lower) modes in (La,Pr)Al$_2$O$_6$. (c) The coupling of $R_4^+$ with $R_5^+$ results in a cooperative lowering of the total energy, stabilizing the non-centrosymmetric structure (grey lines obtained from a fit of the Landau potential, see text). Note that there exists an equivalent energy minimum at $R_5^+\simeq-0.1$ for $R_4^+$ imposed with opposite sense (negative mode amplitude).}
\label{fig:LPAOenergy}
\end{figure*}

LaAlO$_3$ is then ordered with both NdAlO$_3$ and PrAlO$_3$ along the cubic [111] direction,
\footnote{Note that throughout we choose to write the stoichiometric formula in a manner that emphasizes each compounds similarity with double perovskites, but each phase may be equivalently written as an 
ultra-short period 1/1, for example, (La,Nd)Al$_2$O$_6$ is equivalent to an (LaAlO$_3$)$_1$/(NdAlO$_3$)$_1$.}
resulting in two rock salt double perovskites: (La,Nd)Al$_2$O$_6$ and (La,Pr)Al$_2$O$_6$.
CeAlO$_3$ and PrAlO$_3$ were ordered in the same way resulting in a third superlattice: (Ce,Pr)Al$_2$O$_6$.
By employing a symmetry-restricted soft-phonon search, we determined the ground state structures of each compound, which are summarized in \autoref{tab:gs}.
All three ordered aluminates contain only out-of-phase rotations [\autoref{fig:LPAOenergy}(a)], which in combination with the [111]-cation ordering, results in non-centrosymmetric (chiral or polar) space groups.
However, only (La,Pr)Al$_2$O$_6$ and (Ce,Pr)Al$_2$O$_6$ display a spontaneous polarization, as they are polar ($Imm2$), where as (La,Nd)Al$_2$O$_6$ is chiral and non-polar.

We now carry out a detailed examination of each compounds' atomic structure to understand the origin for the non-centrosymmetric symmetries, the different space groups adopted, as well as the microscopic mechanism for the emergence of an electric polarization from the combination of two bulk non-polar dielectrics.

\begingroup
\squeezetable
\begin{table}[t]
\begin{ruledtabular}
\caption{\label{tab:gs}Structural details and ferroelectric properties of the [111]-ordered perovskite aluminates. All compounds exhibit out-of-phase rotations in their ground states, with the rotation magnitude and electric polarization specified by $\Theta$ (degrees) and $\mathcal{P}$ ($\mu$C/cm$^2$), respectively. } 
\begin{tabular}{lccccc}
Compound ($\tau$) & S.G.\ & Tilt &  Lattice Parameters (\AA) & $\Theta$ & $\mathcal{P}$\\
\hline\\[-0.6em]
(La,Nd)Al$_2$O$_6$ & $R32$ & $a^-a^-a^-$ &$a$ = 5.338 & 7.6 & 0 \\
(0.985)  & & & $b$ = 5.338 & & \\
  & &  & $c$ = 12.96 & & \\
\hline
(La,Pr)Al$_2$O$_6$ & $Imm2$ & $a^0b^-b^-$ & $a$ = 7.523 & 6.9 & 1.80 \\
(0.988)  & & & $b$ = 5.355 & & \\
  & & &  $c$ = 5.322 & & \\
\hline
(Ce,Pr)Al$_2$O$_6$ & $Imm2$ & $a^0b^-b^-$ &  $a$ = 7.560 & 6.7 & 1.78 \\
(0.985)&  & & $b$ = 5.399 & & \\
& & &  $c$ = 5.364 & & \\
\end{tabular}
\end{ruledtabular}
\end{table}
\endgroup

\subsubsection{Chiral (La,Nd)Al$_2$O$_6$}

In its ground state, (La,Nd)Al$_2$O$_6$ displays an $a^-a^-a^-$ tilt pattern, which is similar to its two bulk $AB$O$_3$ constituents, and the $R32$ space group.
The distortion relating the undistorted structure to the ground state transforms as a single irreducible representation (irrep), $R_{4}^{+}$, of the undistorted 5-atom $Pm\bar{3}m$, which describes collective oxygen displacements producing out-of-phase rotations along each Cartesian axis.
This may be equivalently described as a single out-of-phase rotation about the trigonal (3-fold) axis of the structure.

The Landau theory of displacive phase transitions allows for the free energy of a system to be expanded in terms of an order parameter; in these superlattices, the order parameter best capturing the symmetry reduction from the cubic to antiferrodistortive phase is the AlO$_6$ octahedral rotation angle (given by $\Theta$ in \autoref{tab:gs}), 
which can be mapped onto  the amplitude of the zone-boundary mode
$R_{4}^{+}$.
This permits us to write a free energy expression for 
(La,Nd)Al$_2$O$_6$ as
 \begin{equation}
 \mathcal{F} = \alpha_1Q_{R_4^+}^{2} + \beta_1Q_{R_4^+}^4\,,
 \label{eqn1}
\end{equation}
with coefficients obtained from fits to our DFT computed mode amplitude--energy plots (\autoref{tab:freeenergy}).

This compound has a null polarization, despite the fact that it exhibits a non-centrosymmetric space group, owing to the chiral structure.
The out-of-phase rotations along all three crystallographic axes (in addition to rock salt ordering) removes all possibility for mirror symmetry but is compatible with three-fold and 2-fold rotation axes. 
As previously reported for hybrid improper ferroelectrics, it is essentially the anti-polar displacements of the different $A$-site cations which results in the electric polarization (a ferr$i$electric type mechanism). \cite{Mulder/Rondinelli/Fennie:2013,Gou/Rondinelli:2013}
However, although out-of-phase rotations alone can lift inversion symmetry in [111]-ordered superlattices, the $A$-site atoms cannot displace perpendicular to any direction which contain out-of-phase rotations. 
These features make the compound chiral and not polar.\footnote{Indeed, in the $R32$ space group, the $A$ and $A^{\prime}$ cations occupy the $3a$ and $3b$ Wyckoff positions, both of which exhibit site symmetry $32$.}
This symmetry constraint suggests that out-of-phase rotations along one or two directions could be compatible with mirror planes in [111] $A$-site ordered perovskites since those directions without any rotations would serve as unique anisotropic axes for which a loci of points defining the reflection plane may exist. 

\subsubsection{Polar (La,Pr)Al$_2$O$_6$ and (Ce,Pr)Al$_2$O$_6$}
We next investigate ordered aluminates which contain out-of-phase rotations along only two crystallographic axes.
Because the bulk phases of the constituents in (La,Pr)Al$_2$O$_6$ and (Ce,Pr)Al$_2$O$_6$ superlattices display different octahedral rotation patterns (\autoref{tab:bulk}), there is a competition between having out-of-phase rotations along one, two, or three axes.
We find that the ground state structures of both $A$-site ordered aluminates exhibit the $a^0b^-b^-$ rotational pattern with the polar $Imm2$ space group.
This rotation pattern allows for displacements of the $A$-sites perpendicular to the $a$ axis, resulting in small electric  polarizations of 1.80\,$\mu$C/cm$^2$ in (La,Pr)Al$_2$O$_6$ and 1.78\,$\mu$C/cm$^2$ in (Ce,Pr)Al$_2$O$_6$.

Unlike (La,Nd)Al$_2$O$_6$, the symmetry of these polar ground state structures requires two irreps for a complete description: $Q_1$, which describes the out-of-phase rotations and transforms like the irrep $R_{4}^{+}$  [identical to that in (La,Nd)Al$_2$O$_6$, \autoref{fig:LPAOenergy}(a)], \emph{and} $Q_2$, which describes $A$-site displacements and transforms like $R_{5}^{+}$ [\emph{cf}.\ inset in \autoref{fig:LPAOenergy}(b)].
Both irreps are given relative to the $Pm\bar{3}m$ 5-atom perovskite.
\autoref{fig:LPAOenergy}(b) shows the evolution of the total energy of  (La,Nd)Al$_2$O$_6$ with respect to increasing amplitude of these two modes.
We find that while $R_{4}^{+}$ is a soft mode (negative quadratic-like curvature about the origin) and results in a large gain,  $R_5^+$ alone leads to an energy penalty (positive curvature).
On the other hand, when the two modes coexist [\autoref{fig:LPAOenergy}(c)], we find that with increasing amplitude of the nominally hard $R_{5}^{+}$ mode any non-zero amplitude of the non-polar out-of-phase rotations ($R_{4}^{+}$) leads to increased stability of the $Imm2$ (La,Nd)Al$_2$O$_6$ structure without a  change in curvature of the energy surface. 
This behavior is similar to that seen in the improper ferroelectric YMnO$_3$,\cite{Fennie/Rabe:2005} whereby a non-polar mode also stabilizes a hard polar mode with non-zero amplitude.
Note, the free energy evolution of (Ce,Pr)Al$_2$O$_6$ exhibits identical behavior (data not shown).

\begingroup
\squeezetable
\begin{table}[b]
\begin{ruledtabular}
\caption{\label{tab:freeenergy}Free energy expansion coefficients to  \autoref{eq:landau} for each  [111]-ordered $A$-site perovskite aluminate.} 
\begin{tabular}{cccc}
Coefficient & (La,Nd)Al$_2$O$_6$ & (La,Pr)Al$_2$O$_6$ & (Ce,Pr)Al$_2$O$_6$ \\
\hline\\[-0.6em]
$\alpha_1$ (meV/\AA$^2$) & -8.130 & -30.61 & -33.98 \\
$\alpha_2$ (meV/\AA$^2$) & -- & 132.9 & 133.1 \\
$\beta_1$ (meV/\AA$^4$) & 0.139 & 2.239 & 2.433 \\
$\beta_2$ (meV/\AA$^4$) & -- & 1.295 & 8.967 \\
$\gamma_1$ (meV/\AA$^4$) & -- & -3.299 & -4.085 \\
$\gamma_2$ (meV/\AA$^4$) & -- & 245.2 & 217.9 \\
$\delta$ (meV/\AA$^4$) & -- & 33.48 & 33.37 \\
\end{tabular}
\end{ruledtabular}
\end{table}
\endgroup

To understand the anharmonic lattice  interactions which provide the stability of the polar phase, we expand the free energy  (with respect to $Pm\bar{3}m$) as
\begin{eqnarray}
\label{eq:landau}
\mathcal{F} &=& \alpha_1Q_{R_4^+}^{2} + \alpha_2Q_{R_5^+}^{2} + \beta_1Q_{R_4^+}^4  +\beta_2Q_{R_5^+}^4\\\nonumber
&+&\gamma_1Q_{R_4^+}^{3}Q_{R_5^+}+\gamma_2Q_{R_4^+}Q_{R_5^+}^3+\delta Q_{R_5^+}^2Q_{R_4^+}^2\,.
\end{eqnarray}
By fitting the free energy expression truncated to quartic order to the calculated total energies in Figure \autoref{fig:LPAOenergy}(c), the coefficients of \autoref{eq:landau} are  obtained (\autoref{tab:freeenergy}).
We find that the most important anharmonic term coupling 
$Q_{R_4^+}$ to $Q_{R_5^+}$ is given by $\gamma_1Q_{R_4^+}^3Q_{R_5^+}$ with a negative coefficient that leads to the cooperative stability and coexistence of the out-of-phase rotations and polar displacements. 
The $\gamma_2Q_{R_4^+}Q_{R_5^+}^3$ term contributes to the asymmetry of the energy surface, but is not the key interaction stabilizing the  non-zero polar displacements in the $Imm2$ structure.
Similarly, the mixed bi-quadratic term renormalizes the homogeneous quadratic terms, and leads to the decrease in radius of curvature of the free energy curve near the minimum with increasing ${R_4^+}$, yet it is not responsible for the minimum.
Thus, the requirement of two coupled modes as $\sim\!Q_{R_4^+}^3Q_{R_5^+}$, with the primary $Q_{R_4^+}$ octahedral rotation mode,  reveals that the aluminates  behave as conventional improper ferroelectrics and not hybrid-improper ferroelectrics; the latter are susceptible to staggered phase transitions owing to the necessity of two primary modes.

\begin{figure*}
\centering
\includegraphics[width=1.65\columnwidth,clip]{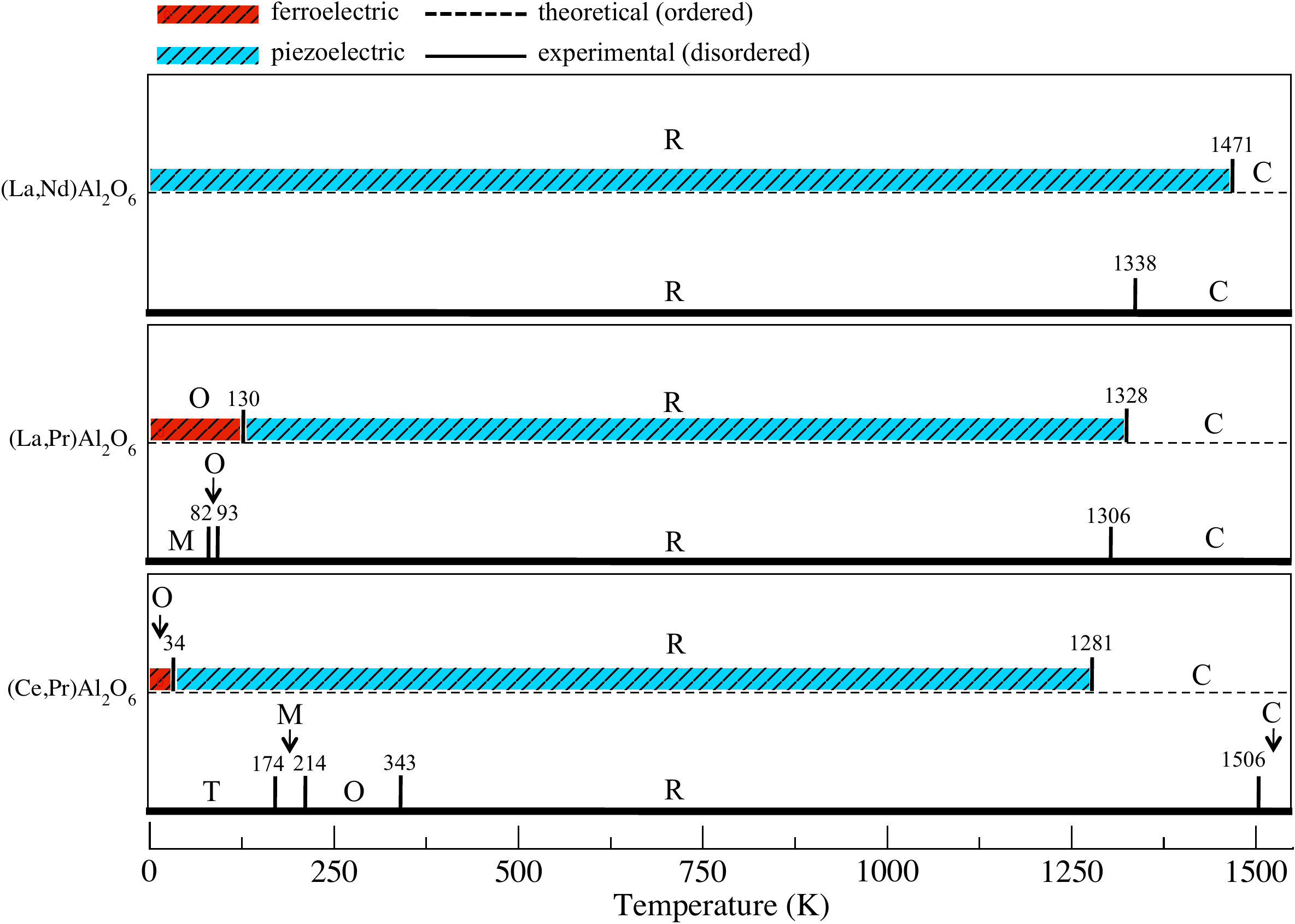}\vspace{-0.8\baselineskip}
\caption{The experimentally and computationally determined phase transitions of (La,Nd)Al$_2$O$_6$, (La,Pr)Al$_2$O$_6$, and (Ce,Pr)Al$_2$O$_6$. The phases are defined according to crystal system, with the space group symmetries specified in \autoref{tab:symmetry}. Note that the unordered solid solutions for each aluminate are centrosymmetric, whereas the  [111] $A$-cation ordering leads to a non-centrosymmetric structure which persist to high temperature ($\ge 1300$ K), allowing for the presence of ferro- and piezoelectricity.}
\label{fig:transitions}
\end{figure*}

\subsection{Structural Phase Transitions}

We next investigate the critical temperatures associated with the paraelectric to ferroelectric transitions to explore the potential of these aluminates for high-temperature applications.
Recent work has shown that $T_c$ in compounds exhibiting soft-mode driven phase transitions is strongly correlated to the energy difference between the two symmetry related structures.\cite{Wojdel/Iniguez:2013}
By using the fact that $\Delta E=k_BT_c$, where $k_B$ is Boltzmann's constant, we are able to estimate phase transition temperatures using DFT total energies.
The phase diagrams of the rare earth aluminate solid solutions (effectively disordered $A$-site phases) of all compounds investigated in this work have also been experimentally determined,\cite{Vasylechko2009113} and thus will serve as a useful comparison to the predicted critical temperatures of the  [111] ordered phases.
Because the symmetries of an ordered and disordered compound with the same tilt pattern are different, we refer to the corresponding phases by the crystal system generated by the octahedral rotation pattern to more easily draw comparisons.
The abbreviations used for each phase, along with the corresponding tilt pattern and space group are listed in \autoref{tab:symmetry}.

\begingroup
\squeezetable
\begin{table}[t]
\begin{ruledtabular}
\caption{\label{tab:symmetry}The possible crystal systems, octahedral rotation patterns, and space groups (S.G.) that may be found in (La,Nd)Al$_2$O$_6$, (La,Pr)Al$_2$O$_6$, or (Ce,Pr)Al$_2$O$_6$, along with the abbreviations used for each. Note that not every aluminate displays every phase tabulated below (see \autoref{fig:transitions}).} 
\begin{tabular}{llll}
Crystal System & Tilt Pattern & Ordered S.G. & Solid Solution S.G. \\
\hline\\[-0.6em]
(T) Tetragonal & $a^0a^0c^-$ & $I\bar{4}2m$ (121) & $I4/mcm$ (140) \\
(M) Monoclinic & $a^0b^-c^-$ & $Cm$ (8) & $C2/m$ (12) \\
(O) Orthorhombic & $a^0b^-b^-$ & $Imm2$ (31) & $Pnma$ (62) \\
(R) Rhombohedral & $a^-a^-a^-$ & $R32$ (155) & $R\bar{3}c$ (167) \\
(C) Cubic & $a^0a^0a^0$ & $Fm\bar{3}m$ (225) & $Pm\bar{3}m$ (221) \\
\end{tabular}
\end{ruledtabular}
\end{table}
\endgroup

Experimentally, solid solution (La,Nd)Al$_2$O$_6$ is found to undergo a rhombohedral (R, $a^-a^-a^-$) to cubic (C, $a^0a^0a^0$) phase transition at 1471 K.
Our calculations of the ordered [111]-system estimate this transition to occur at 1338 K, in  good agreement with the experimental results (\autoref{fig:transitions}, top panel).\footnote{%
Note that other ordered phases (such as [001]) could display different transition temperatures.}

Solid solution (La,Pr)Al$_2$O$_6$ displays a  more complicated series of phase transitions: 
\[
\mathrm{M} \xrightarrow{82 \: \textup{\scriptsize K}} \mathrm{O} \xrightarrow{93 \: \textup{\scriptsize K}} \mathrm{R} \xrightarrow{1306 \: \textup{\scriptsize K}} \mathrm{C}.
\]
The monoclinic (M) phase exhibits the $a^0b^-c^-$ tilt pattern, while the orthorhombic (O) phase exhibits the $a^0b^-b^-$ pattern.
In our calculations of the [111]-ordered phase, however, the monoclinic $Cm$ structure was dynamically unstable to the orthorhombic $Imm2$ phase; no local minimum could be found. Thus the experimentally known sequence of phase transitions in the solid solution are not  reproduced in the ordered phase  (\autoref{fig:transitions}, center panel). This result may either be a consequence of the cation order or a limitation of the  exchange-correlation functional used in the DFT calculations.
However, we estimate the $\mathrm{O} \rightarrow \mathrm{R}$ and $\mathrm{R} \rightarrow \mathrm{C}$ transitions to occur at 130 K and 1328 K, respectively, in good agreement with the experimental results of 93 K and 1306 K, which suggests that the monoclinic phase is suppressed by the cation order.
Additionally, we found an orthorhombic $Pmc2_1$ ($a^+b^-b^-$) that is $\sim$20 meV higher in energy than the ground state $Imm2$ structure; although this phase is not reported experimentally in the solid solution phase diagram, it could manifest during growth of [111]-ordered (La,Pr)Al$_2$O$_6$.

Finally, solid solution (Ce,Pr)Al$_2$O$_6$ displays the most complicated phase diagram, having five  experimental transitions:\cite{Vasylechko20071277}
\[
\mathrm{T} \xrightarrow{174 \: \textup{\scriptsize K}} \mathrm{M} \xrightarrow{214 \: \textup{\scriptsize K}} \mathrm{O} 
\xrightarrow{343 \: \textup{\scriptsize K}}\mathrm{R} \xrightarrow{1506 \: \textup{\scriptsize K}} \mathrm{C}.
\]
In contrast, we find the orthorhombic $Imm2$ structure to be the ground state phase and the tetragonal structure to be unstable (\autoref{fig:transitions}).
Based on energetics, we estimate an $\mathrm{O} \rightarrow \mathrm{R}$ transition at 34 K and an $\mathrm{R} \rightarrow \mathrm{C}$ transition at 1281 K.
We conjecture that many of the discrepancies between the ordered aluminates and the experimental solid solutions, \emph{e.g.}, the loss of the tetragonal phase in (Ce,Pr)Al$_2$O$_6$,  may be attributed to the $A$-site rock salt ordering pattern and its compatibility with the single antiferrodistortive AlO$_6$ rotation mode.
Because ordering along [111] results in a three dimensional pattern (alternating $A$-sites along each axis), it is more compatible with a three dimensional octahedral rotation pattern ($a^-a^-a^-$, \textit{i.e.}, out-of-phase rotations along each axis), rather than a one- or two-dimensional tilt pattern which would create a unique axis in the structure.
This geometric argument is corroborated by examining rock salt $B$-site ordered perovskites: while rock salt $A$-site ordered perovskite oxides are rare, [111]-$B$-site ordering is by far the most common type.\cite{Poeppelmeier:1993,Graham/Woodward:2010,Davies_2008}
The three-dimensional $a^+b^-b^-$ tilt system is the most common octahedral rotation pattern adopted by $B$-site ordered perovskites (in addition to being the most common overall). The $a^-a^-a^-$ and $a^0b^-b^-$ patterns are found to be the next most frequently observed, with $a^0a^0c^-$ being the least common.\cite{Poeppelmeier:1993,Howard/Kennedy/Woodward:2003}
By considering this distribution of structures, it seems reasonable that the  ground states of the $A$-site ordered aluminates investigated here should also exhibit either the $a^-a^-a^-$ or $a^0b^-b^-$ rotational pattern.

Additionally, it appears that PrAlO$_3$ ($a^0b^-b^-$) influences the tilt pattern of the ordered superlattice more so than the other bulk rare earth aluminates; it turns off a rotation about one Cartesian axis in  LaAlO$_3$ ($a^-a^-a^-$), stabilizing the  $a^0b^-b^-$ rotation pattern.
Whereas it turns on a rotation when ordered with CeAlO$_3$ ($a^0a^0c^-$) to also give  the $a^0b^-b^-$ rotation pattern.
This is due to the fact that the phonon mode which transforms as $R_4^+$  is much more unstable in PrAlO$_3$ ($\omega=177.5i$ cm$^{-1}$) than in LaAlO$_3$ ($\omega=127.1i$ cm$^{-1}$) or CeAlO$_3$ ($\omega=148.4i$ cm$^{-1}$), leading to control of the superlattice structure by PrAlO$_3$.
This suggests that the rotation pattern adopted by the ground state structure could be directly designed by selection  of the rotational mode instability strengths (or mismatch) of the bulk $AB$O$_3$ perovskite oxides interleaved to form the superlattice.

\begingroup
\squeezetable
\begin{table*}[t]
\begin{ruledtabular}
\caption{\label{tab:piezo}The computed frozen-ion ($\epsilon^{elec}$), relaxed-ion ($\epsilon^{ion}$), and total ($\epsilon$) dielectric,  relaxed-ion piezoelectric stress ($e$) and strain ($d$), and Born effective charge ($Z^*$) tensors for (La,Nd)Al$_2$O$_6$, (La,Pr)Al$_2$O$_6$, and (Ce,Pr)Al$_2$O$_6$. Because (La,Nd)Al$_2$O$_6$ does not exhibit any other phases besides $R32$ and $Fm\bar{3}m$, only the tensor properties of the  $R32$ phase are presented. Point group $32$ exhibits 3 dielectric and 5 piezoelectric coefficients, but only 2 of which are independent in each case: $\epsilon_{11}=\epsilon_{22}\neq\epsilon_{33}$, and $d_{11}=-d_{12}=-\frac{1}{2}d_{26}$ and $d_{14}=-d_{25}$. The $mm2$ point group exhibits 3 independent dielectric and 5 independent piezoelectric coefficients. Although the Al and O Born effective charge tensors contain small off-diagonal elements, only the main diagonal coefficients are given.} 
\begin{tabular}{cccccccccccccc}
& & \multicolumn{4}{c}{Dielectric Constant} & \multicolumn{3}{c}{Piezoelectric Coefficients} & \multicolumn{5}{c}{Born Effective Charges} \\
\cline{3-6}
\cline{7-9}
\cline{10-14}
Compound & Phase & Index & $\epsilon^{elec}$ & $\epsilon^{ion}$ & $\epsilon$ & Index & $e$ (C/m$^2$) & $d$ (pC/N) & Atom & nominal & $Z^*_{11}$ & $Z^*_{22}$ & $Z^*_{33}$ \\
\hline\\[-0.6em]
(La,Nd)Al$_2$O$_6$ & $R32$ & 11 & 4.71 & 49.7 & 54.4 &11 & 6.83 & 13.3 & La & $+3$ & $+4.47$ & $+4.47$ & $+4.21$ \\
 & & 33 & 4.61 & 27.6 & 32.2 & 15 & 4.08 & 77.7 & Nd & $+3$ & $+4.31$ & $+4.31$ & $+4.14$ \\
 & & & & & & 26 & -6.83 & -26.6 & Al & $+3$ & $+2.95$ & $+2.95$ & $+2.86$ \\
 & & & & & & & & &  O & $-2$ & $-2.42$ & $-2.42$ & $-2.38$ \\
 \hline
(La,Pr)Al$_2$O$_6$ & $Imm2$ & 11 & 4.70 & 23.6 & 28.3 & 31 & 0.11 & 0.11 & La & $+3$ & $+4.38$ & $+4.25$ & $+4.58$ \\
 & & 22 & 4.63 & 18.8 & 23.4 & 32 & 0.19 & 0.62 & Pr & $+3$ & $+4.30$ & $+4.14$ & $+4.44$ \\
 & & 33 & 4.78 & 17.4 & 22.2 & 33 & 0.02 & -0.19 & Al & $+3$ & $+2.88$ & $+2.94$ & $+2.95$ \\
 & & & & & & 24 & 0.18 &  9.05 & O & $-2$ & $-2.42$ & $-2.42$ & $-2.38$ \\
 & & & & & &15 & -0.19 & -1.24 & & & & &\\ 
\cline{2-14}
 & $R32$ & 11 & 4.74 & 36.1 & 40.8 & 11 & 6.83 & 5.23 & La & $+3$ & $+4.72$ & $+4.72$ & $+4.26$ \\
 & & 33 & 4.65 & 25.9 & 30.5 & 15 & 3.28 & 54.4 & Pr & $+3$ & $+4.35$ & $+4.35$ & $+4.20$ \\
 & & & & & &  26 & -6.83 & -10.5 & Al & $+3$ & $+2.95$ & $+2.95$ & $+2.88$ \\
 & & & & & & & & &  O & $-2$ & $-2.44$ & $-2.44$ & $-2.39$ \\
 \hline
(Ce,Pr)Al$_2$O$_6$ & $Imm2$ & 11 & 4.67 & 22.4 & 27.1 & 31 & 0.07 & 0.14 & Ce & $+3$ & $+4.31$ & $+4.14$ & $+4.47$ \\
 & & 22 & 4.58 & 17.6 & 22.2 & 32 & 0.05 & 0.04 & Pr & $+3$ & $+4.30$ & $+4.13$ & $+4.44$ \\
 & & 33 & 4.75 & 16.1 & 20.8 & 33 & 0.06 & 0.12 & Al & $+3$ & $+2.88$ & $+2.95$ & $+2.96$ \\
 & & & & & & 24 & 0.09 & -3.39 &  O & $-2$ & $-2.43$ & $-2.35$ & $-2.47$ \\
 & & & & & & 15 & -0.11 & -0.71 \\ 
\cline{2-14}
& $R32$ & 11 & 4.70 & 17.3 & 22.0 & 11 & 0.46 & 2.25 & Ce & $+3$ & $+4.36$ & $+4.36$ & $+4.18$ \\
 & & 33 & 4.61 & 25.0 & 29.6 & 15 & 0.20 & 0.26 & Pr & $+3$ & $+4.34$ & $+4.134$ & $+4.18$ \\
 & & & & & & 26 & -0.46 & -4.49 & Al & $+3$ & $+2.96$ & $+2.96$ & $+2.87$ \\
  & & & & & & & & &  O & $-2$ & $-2.44$ & $-2.44$ & $-2.36$ \\
\end{tabular}
\end{ruledtabular}
\end{table*}
\endgroup

\subsection{Dielectric and Acentric Properties}

As discussed previously, we find the ground state structures of  (La,Pr)Al$_2$O$_6$ and (Ce,Pr)Al$_2$O$_6$ are improper ferroelectrics with small electric polarizations of 1.80 and 1.78 $\mu$C/cm$^2$, respectively, resulting from $A$-site displacements.
Our estimate of the critical transition temperatures for each shows that these aluminates would only retain this spontaneous polarization to 93 K and 34 K, at which point they transition from a polar to chiral structure.
However, all three compounds retain non-centrosymmetric structures up to high temperatures (\textgreater 1300 K, see \autoref{fig:transitions}).
To investigate their viability for use in high-temperature  applications, we compute the dielectric and piezoelectric tensors, as well as the Born effective charges, for the non-centrosymmetric phases of each ordered compound (\autoref{tab:piezo}).
\footnote{Although these properties are all computed at 0 K, they can still be used to draw useful comparisons across different phases.}
The total dielectric constant $\epsilon$ consists of electronic (frozen-ion) and ionic (relaxed-ion) contributions.
We first find that the electronic dielectric tensor is approximately isotropic, as well as nearly equivalent across all chemical compositions (a well-known phenomenon in perovskite oxides\cite{dielectric_perovs_2010})
 and structural phases.
The ionic contributions are much larger; we also find that on average they increase when transitioning from the polar $Imm2$ to chiral $R32$ structure. 
The total dielectric constant of each aluminate also falls well within the known range for perovskite oxides; (La,Nd)Al$_2$O$_6$ in particular displays a dielectric constant near the top of this range.\cite{dielectric_perovs_2010}

The relaxed-ion piezoelectric stress ($e_{ij}$) and strain ($d_{ij}$) coefficients are related through the compliance tensor $C_{ij}$ (the inverse of the elastic tensor) by $d_{ij}=S_{ik}e_{kj}$.
Note that due to symmetry, the $R32$ piezoelectric tensors have 5 components, but only 2 of which are independent; the $Imm2$ phase has 5 independent coefficients.
First, the chiral $R32$ phases of (La,Nd)Al$_2$O$_6$ and (La,Pr)Al$_2$O$_6$ exhibit relatively large piezoelectric coefficients, comparable to those of common lead-free piezoelectric materials such as BaTiO$_3$ and LiNbO$_3$.\cite{Kong/Zhang/Ma/Boey:2008,Ledbetter/Ogi/Nakamura:2004}
Interestingly, the \emph{polar} $Imm2$ phase of (La,Pr)Al$_2$O$_6$ has piezoelectric coefficients that are an order of magnitude smaller.
Because the chiral phases have no net dipole due to the $A$-sites sitting on high symmetry positions, an applied stress will generate a much larger induced polarization than in the polar phases (which already have a spontaneous polarization).
Additionally, the piezoelectric response of (Ce,Pr)Al$_2$O$_6$ is much smaller than either of the La-based aluminates; it does, however, also show the same increase in response across the $Imm2$-to-$R32$ structural phase boundary (\autoref{tab:piezo}).
Thus, while the polarization goes to zero across the $Imm2$-to-$R32$ transition in (La,Pr)Al$_2$O$_6$ and (Ce,Pr)Al$_2$O$_6$, there is a large increase in the piezoelectric response in the chiral structures.

Finally, the Born effective charges were computed for the five phases.
Typically, proper $AB$O$_3$ ferroelectrics exhibit anomalously large Born effective charges; for example, the nominal charge for Ba, Ti, and O in BaTiO$_3$ are $+2$, $+4$, and $-2$, respectively, while $Z_\mathrm{Ba}^* = +2.56$, $Z_\mathrm{Ti}^* = +7.26$, and $Z^*_\mathrm{O} = -5.73$.\cite{King-Smith/Vanderbilt-1994}
In these ordered aluminates, however, the Born effective charges are close to their nominal values, with only the $A$-site cations displaying any significant deviation.
This clearly highlights the difference in mechanism between proper ferroelectrics such as BaTiO$_3$, where polarization arises from $B$-site displacements due to enhanced covalency with the surrounding oxygen anions, and improper ferroelectrics such as these ordered aluminates, where unequal anti-aligned $A$-site displacements result in layer dipoles.

\section{Conclusions}
We showed that the ordering of $A$-site cations along the [111]-crystallographic axis in perovskite oxide superlattices, in conjunction with any rotational pattern containing only out-of-phase rotations results in a non-centrosymmetric space group.
By selecting compounds which contain ground state rotational patterns described by only one lattice mode (irreducible representation $R_4^+$), we were able to design three rock salt ordered superlattices, two of which display a spontaneous electric polarization originating from an improper ferroelectric mechanism.
This type of design strategy prevents staggered paraelectric to ferroelectric phase transitions, possible in hybrid improper ferroelectrics where the most frequently pursued design route requires the $a^+b^-b^-$ tilt pattern, without sacrificing the desired electronic properties.
Additionally, we predicted the phase transition temperatures of each superlattice from first-principles, which showed that these compounds remain non-centrosymmetric up to 1300 K.
Across the ferroelectric phase transition, there is an order of magnitude increase in the piezoelectric coefficients comparable to conventional piezoelectric materials such as BaTiO$_3$.
This indicates that these materials may find potential use in high-temperature applications, as they have reasonable piezoelectric responses  that persist over a broad temperature range.
We hope the framework presented here encourages the synthesis of ultra-short perovskite superlattices on [111]-oriented substrates, and that it can be extended beyond these aluminates to other perovskite oxides with out-of-phase rotations.

\begin{acknowledgments}
J.Y.\ and J.M.R.\ acknowledge funding support from the Army Research Office under grant number  W911NF-12-1-0133. J.Y. thanks members of the Rondinelli group for useful discussions. 
DFT calculations were performed using the CARBON cluster at the Center for Nanoscale Materials  
[Argonne National Laboratory, supported by the U.S.\ DOE, Office of Basic Energy Sciences (BES), DE-AC02-06CH11357], and the 
Extreme Science and Engineering Discovery Environment (XSEDE), which is supported by National Science Foundation grant number OCI-1053575.
\end{acknowledgments}

%

\end{document}